\documentclass[prd,twocolumn,aps,showpacs,preprintnumbers,amsmath,amssymb]{revtex4}
\usepackage{graphicx}
\usepackage{dcolumn}
\usepackage{bm}
\usepackage{psfrag}
\usepackage{here}
\begin{document}
\newcommand{\be}{\begin{equation}}\newcommand{\ee}{\end{equation}}
\newcommand{\bea}{\begin{eqnarray}}\newcommand{\eea}{\end{eqnarray}}
\newcommand{\bc}{\begin{center}}\newcommand{\ec}{\end{center}}
\def\no{\nonumber}
\def\eq#1{Eq. (\ref{#1})}\def\eqeq#1#2{Eqs. (\ref{#1}) and  (\ref{#2})}
\def\lsim{\raise0.3ex\hbox{$\;<$\kern-0.75em\raise-1.1ex\hbox{$\sim\;$}}}
\def\gsim{\raise0.3ex\hbox{$\;>$\kern-0.75em\raise-1.1ex\hbox{$\sim\;$}}}
\def\slash#1{\ooalign{\hfil/\hfil\crcr$#1$}}\def\eff{\mbox{\tiny{eff}}}
\def\order#1{{\mathcal{O}}(#1)}
\def\pppm{B^0\to\pi^+\pi^-}
\def\pzpz{B^0\to\pi^0\pi^0}
\def\pppz{B^0\to\pi^+\pi^0}
\preprint{ }
\title{Testing QCDF factorization with phase determinations in $B\to K\pi$,$K\rho$, and
 $ K^{*}\pi$ decays}
\author{
T. N. Pham$^{1}$\footnote{email address: pham@cpht.polytechnique.fr}}
\affiliation{$^1$ Centre de Physique Th\'{e}orique, CNRS, Universit\' e  Paris-Saclay\\ 
Ecole Polytechnique, 91128 Palaiseau, Cedex, France }
\date{\today}
\begin{abstract}
The success of QCD factorization(QCDF)  in predicting branching
ratios for charmless $B$ decays to light pseudoscalar and  vector 
mesons  and the  small CP asymmetries measured at
 $BABAR$, Belle and LHCb show that the  phase in these decays, as predicted 
by QCDF, are not large. For a  precise test of QCDF
one needs to extract from the measured  decay rates,  the phase of  
the decay amplitude which appears  in the interference terms  between the 
tree  and penguin contribution. Since the tree amplitude is 
known at the leading order in $\Lambda_{\rm QCD}/m_{b}$ and is consistent with
the  measured tree-dominated decay rates, the QCDF value for the tree amplitude
can be used with the measured  decay rates to obtain the
phases in $B\to K\pi$,$K\rho$, and  $ K^{*}\pi$ decay rates. This is similar
to the extraction   of the final-state interaction phases in
the interference term  between  $p\bar{p}\to J/\Psi\to e^{+}e^{-}$
and $p\bar{p}\to e^{+}e^{-} $  and in $J/\Psi\to 0^{-}0^{-}$ 
done previously. In this paper, we present a determination of the phase
 between the $I=3/2$ tree and $I=1/2$  penguin amplitudes in 
$B\to K\pi$,$K\rho$ , and  $ K^{*}\pi$ decays using the measured decay 
rates and the QCDF $I=3/2$ tree amplitude obtained from the $I=2$ 
$B^{+}\to \pi^{+}\pi^{0},\rho^{0}\pi^{+}, \rho^{+}\pi^{0}$
tree-dominated decays   and   compare the result with the phase 
given by QCDF. It is remarkable that the phase extracted from 
experiments differs only slightly from the QCDF values. This shows 
that there is no large final-state interaction  strong phase  in 
$B\to K\pi$,$K\rho$, and  $ K^{*}\pi$ decays.
\end{abstract}
\maketitle
\section{INTRODUCTION}
QCD Factorization(QCDF)\cite{Beneke,Beneke2} seems to be rather successful in 
predicting branching ratios and CP asymmetries for charmless $B$ decays
into light pseudoscalar and  vector mesons. The small CP asymmetries
measured at {\it BABAR},Belle and LHCb show that the final-state interaction
phase in these decays, as predicted by QCDF, is not large. For 
penguin-dominated  charmless $B$ decays into two light pseudoscalar
and vector mesons,  the phase appearing in the decay amplitude is the
relative phase between the isospin $I=3/2$  tree and $I=1/2$  penguin 
amplitude, as in $B\to K\pi$,$K\rho$ , and  $ K^{*}\pi$ 
decays. Since all four modes for $B\to K\pi$,  $K\rho$, and
 $ K^{*}\pi$,   respectively, have similar branching ratios, the
interference terms are quite small, making a determination of
these phases more difficult  than for the Cabibbo-favored decays
 $D\to \bar{K\pi}$,$\bar{K}\rho$, and  $ \bar{K}^{*}\pi$, for which a large 
$\delta^{K\pi}_{3/2}-\delta^{K\pi}_{1/2}= (86\pm 8^{\circ})$ has been 
obtained\cite{Kamal}.  Since the tree amplitude is 
known at the leading order in $\Lambda_{\rm QCD}/m_{b}$ \cite{Beneke2} 
and is consistent with
the  measured tree-dominated decay rates, knowledge of the tree
amplitude then allows a simple determination of the phase in the decay 
amplitude using the measured decay rates. This is similar
to the extraction   of the final-state interaction phases in
the interference term  between  $p\bar{p}\to J/\Psi\to e^{+}e^{-}$
and $p\bar{p}\to e^{+}e^{-} $ \cite{Baldini} and 
in the process $J/\Psi\to 0^{-}0^{-}$ via three-gluon and one-photon
exchange interference terms \cite{Suzuki}. By expressing the
$B\to PP, PV$ decay amplitudes in terms of the $I=1/2$ and $I=3/2$
isospin amplitudes\cite{Deshpande,Isola}, the relative phase of the two
isospin amplitudes can be obtained from the  magnitudes of the isospin 
amplitudes  and the decay rates, as  knowledge of the three sides of the
triangle formed with the decay amplitude and the other two sides, the two
isospin amplitudes, allows a  determination of the three angles of
the triangle and the corresponding relative phases of the 
amplitudes. This is possible for the penguin-dominated $\Delta S=1$,
 $B\to PP, PV$ decays for which all  the decay rates have been 
measured, and since  QCDF predictions for the $I=2$
 $B^{+}\to \pi^{+}\pi^{0},\rho^{0}\pi^{+}, \rho^{+}\pi^{0}$ 
tree-dominated decays  agree rather well with experiments
as shown in the table below  and  in \cite{Cheng}, the  $I=2$  amplitudes
in these decays  could  be taken as  the $I=3/2$ tree amplitudes in  
penguin-dominated $B\to PP,PV$ decays with $SU(3)$ 
breaking effects in  the $B\to K,K^{*}$ form factors  and decay
constants  involving $K,K^{*}$ meson taken into  account\cite{Ali}. With 
the $I=3/2$ tree amplitude known, the three sides of the triangle formed 
with the decay rate,  the $I=1/2$ and $I=3/2$ isospin
amplitude allows a determination of the three angles and the relative phase
between the sides. In this paper we will present a determination
of the relative phase between the  $I=3/2$ and $I=1/2$ amplitudes 
using  the QCDF  $I=3/2$ amplitude and the measured decay rates. It is
remarkable that the phase extracted from experiments differs
only slightly from the QCDF values.  This shows that  
final-state interaction phases are not large in charmless in $\Delta S=1$
$B\to PP, PV$ decays.  In  the following section we give  amplitudes
and branching ratios  for  $B\to K\pi$,$K\rho$, and $ K^{*}\pi$
decays in the QCD  factorization approach. The  determination of the
phases of the decay  amplitudes obtained from the measured decay rates 
and  from the  QCDF  amplitudes and decay rates are  given in  Sec. III. 

\section{$\Delta S=1$ $B\to PP,PV$   DECAY IN QCD FACTORIZATION}
The $B\to M_{1} M_{2}$, decay amplitude in QCDF for $B=B^{-},\bar{B}^{0}$ 
 is given by \cite{QCDF1,QCDF2}:
\bea
&&  {\cal A}(B \rightarrow M_1 M_2)=
 \frac{G_F}{\sqrt{2}}\sum_{p=u,c}V_{pb}V^{*}_{ps}\times  \nonumber \\
&& \left( -\sum_{i=1}^{10} a_i^p
   \langle M_1 M_2 \vert O_i \vert B \rangle_H + 
 \sum_{i}^{10} f_B f_{M_1}f_{M_2} b_i \right ),
\label{BMM}
\eea 
where the QCD coefficients  $a_{i}^{p}$ contain the vertex corrections,
penguin corrections, and hard spectator scattering contributions, 
the hadronic matrix elements $ \langle M_1 M_2 \vert O_i \vert B
\rangle_H $  of the tree and penguin operators $O_{i}$ are given 
by the factorization model \cite{Ali,Zhu2}, and  $b_{i}$ are the annihilation
 terms. The values for $a_{i}^{p}$,$p=u,c$ , computed from 
the expressions in \cite{QCDF1,QCDF2} at the renormalization 
scale $\mu=m_{b}$, with $m_{b}=4.2\,\rm GeV$, are~:
\bea
&& a_{4}^{c}=-0.031 - 0.010\,i + 0.0009\,\rho_{H}\exp(i\phi_{H}),\nonumber \\
&& a_{4}^{u}=-0.027 - 0.017\,i + 0.0009\,\rho_{H}\exp(i\phi_{H}),\nonumber \\
&& a_{6}^{c}=-0.045 - 0.003\,i ,\quad a_{6}^{u}=-0.042 - 0.013\,i ,\nonumber \\
&& a_{8}^{c}=-0.0004 - 0.0001\,i ,\quad  a_{8}^{u}= 0.0004 - 0.0001\,i ,\nonumber \\
&& a_{10}^{c}=-0.0011 - 0.0001\,i - 0.0006\,\rho_{H}\exp(i\phi_{H}) ,\nonumber \\
&& a_{10}^{u}=-0.0011 + 0.0006\,i - 0.0006\,\rho_{H}\exp(i\phi_{H}).
\label{aiuc}
\eea
for $i=4,6,8,10$. For other coefficients, $a_{i}^{u}=a_{i}^{p}=a_{i}$ :
\bea
&& a_{1}= 1.02 + 0.015\,i -0.012\,\rho_{H}\exp(i\phi_{H}),\nonumber \\
&& a_{2}= 0.156 - 0.089\,i + 0.074\,\rho_{H}\exp(i\phi_{H}), \nonumber \\
&& a_{3}= 0.0025 + 0.0030\,i - 0.0024\,\rho_{H}\exp(i\phi_{H}), \nonumber \\
&& a_{5}=-0.0016 - 0.0034\,i + 0.0029\,\rho_{H}\exp(i\phi_{H}), \nonumber \\
&& a_{7} =-0.00003 - 0.00004\,i - 0.00003\,\rho_{H}\exp(i\phi_{H})\nonumber \\
&& a_{9} = -0.009 - 0.0001\,i + 0.0001\,\rho_{H}\exp(i\phi_{H}).
\label{ai}
\eea
where the complex parameter $\rho_{H}\exp(i\phi_{H})$ represents the 
end-point singularity term in the hard-scattering
corrections $X_{H}=(1
+\rho_{H}\exp(i\phi_{H}))\,\ln(\frac{m_{B}}{\Lambda_{h}})$ \cite{QCDF1,QCDF2}. 

For the annihilation terms, for $B\to PP$  decays, we have~:
\bea
 b_{2}\kern -0.2cm&=&\kern -0.2cm -0.0041 -\kern -0.1cm 0.0071\rho_{A}\exp(i\phi_{A}) - 0.0019(\rho_{A}\exp(i\phi_{A}))^{2} ,\nonumber \\
 b_{3}\kern -0.2cm&=& \kern -0.2cm -0.0071 -\kern -0.1cm 0.016\rho_{A}\exp(i\phi_{A}) - 0.0093(\rho_{A}\exp(i\phi_{A}))^{2}, \nonumber \\
 b_{3}^{ew}\kern -0.2cm&=&\kern -0.2cm -0.00012  -0.00016\,
\rho_{A}\exp(i\phi_{A})\nonumber \\
 &+&\kern -0.2cm 0.000003\,(\rho_{A}\exp(i\phi_{A}))^{2}.
\label{b3}
\eea
where $b_{i}$ are evaluated with the factor $f_{B}f_{M_{1}}f_{M_{2}}$
included and normalized relative to the factor
$f_{K}F^{B\pi}_{0}(m_{B}^{2}-m_{\pi}^{2})$ in the factorizable terms,
and $\rho_{A}$ , like $\rho_{H}$, appears in the divergent 
annihilation term 
$X_{A}=(1 +\rho_{A}\exp(i\phi_{A}))\,\ln(\frac{m_{B}}{\Lambda_{h}})$.

The  $B\to K\pi$ decay amplitude with the factorizable part \cite{Ali}
and the annihilation term \cite{QCDF1,QCDF2,Zhu3} is:
\bea
\kern -0.6cm&&A(B^{+}\to K^{+}\pi^{0}) = -i\frac{G_{F}}{2}f_{K}F^{B\pi}_{0}(m^{2}_{K})
(m_{B}^{2}-m_{\pi}^{2})\nonumber \\
\kern -0.6cm&&\left(V_{ub}V^{*}_{us}a_{1}
+(V_{ub}V^{*}_{us}+V_{cb}V^{*}_{cs})[a_{4} + a_{10} + (a_{6}+a_{8})r_{\chi}]\right)\nonumber \\ 
\kern -0.6cm&& -i\frac{G_{F}}{2}f_{\pi}F^{BK}_{0}(m^{2}_{\pi})(m_{B}^{2}-m_{K}^{2})\nonumber \\ 
\kern -0.6cm
&&\times\left(V_{ub}V^{*}_{us}a_{2}+(V_{ub}V^{*}_{us}+V_{cb}V^{*}_{cs})\times
  \frac{3}{2}(a_{9}-a_{7})\right)\nonumber \\ 
\kern -0.6cm&&-i\frac{G_{F}}{2}f_{B}f_{K}f_{\pi}\nonumber \\ 
\kern -0.6cm&&\times\left[V_{ub}V^{*}_{us}b_{2} 
+ (V_{ub}V^{*}_{us}+V_{cb}V^{*}_{cs})\times(b_{3} + b_{3}^{ew})\right]
\label{K1p0} \\
\kern -0.6cm&&A(B^{+}\to  K^{0}\pi^{+}) =-i\frac{G_{F}}{\sqrt{2}}f_{K}F^{B\pi}_{0}(m^{2}_{K})(m_{B}^{2}-m_{\pi}^{2})\nonumber \\
\kern -0.6cm&&\times(V_{ub}V^{*}_{us}+V_{cb}V^{*}_{cs})\left[a_{4} - \frac{1}{2}a_{10} + (a_{6}-\frac{1}{2}a_{8})r_{\chi}\right]\nonumber \\
\kern -0.6cm&&-i\frac{G_{F}}{\sqrt{2}}f_{B}f_{K}f_{\pi}\nonumber \\ 
\kern -0.6cm&&\times\left[V_{ub}V^{*}_{us}b_{2} 
+ (V_{ub}V^{*}_{us}+V_{cb}V^{*}_{cs})\times(b_{3} + b_{3}^{ew})\right]
\label{K0p1}
\eea
and for $B^{0}$ :
\bea
\kern -0.6cm&&A( B^{0}\to K^{+}\pi^{-}) = -i\frac{G_{F}}{\sqrt{2}}f_{K}F^{B\pi}_{0}(m^{2}_{K})(m_{B}^{2}-m_{\pi}^{2})\nonumber \\
\kern -0.6cm&&\biggl(\kern -0.1cm V_{ub}V^{*}_{us}a_{1}+\kern -0.1cm (V_{ub}V^{*}_{us}+\kern -0.1cm
  V_{cb}V^{*}_{cs})[a_{4} +\kern -0.1cm a_{10} + \kern
-0.1cm(a_{6}+a_{8})r_{\chi}]\kern -0.1cm\biggr)\nonumber\\
\kern -0.6cm&&-i\frac{G_{F}}{\sqrt{2}}f_{B}f_{K}f_{\pi}\left[
 (V_{ub}V^{*}_{us}+V_{cb}V^{*}_{cs})\times(b_{3} - \frac{b_{3}^{ew}}{2})\right]
\label{K1p2}\\
\kern -0.6cm&&A( B^{0}\to  K^{0}\pi^{0}) =i\frac{G_{F}}{2}f_{K}F^{B\pi}_{0}(m^{2}_{K})(m_{B}^{2}-m_{\pi}^{2})\nonumber \\
\kern -0.6cm&&\times(V_{ub}V^{*}_{us}+V_{cb}V^{*}_{cs})\left[a_{4} - \frac{1}{2}a_{10} + (a_{6}-\frac{1}{2}a_{8})r_{\chi}\right]\nonumber \\
\kern -0.6cm&& -i\frac{G_{F}}{2}f_{\pi}F^{BK}_{0}(m^{2}_{\pi})(m_{B}^{2}-m_{K}^{2})\nonumber\\ 
\kern -0.6cm&&\left(V_{ub}V^{*}_{us}a_{2}+(V_{ub}V^{*}_{us}+V_{cb}V^{*}_{cs})\times \frac{3}{2}(a_{9}-a_{7})\right)\nonumber \\ 
\kern -0.6cm&&+i\frac{G_{F}}{2}f_{B}f_{K}f_{\pi}\left[
 (V_{ub}V^{*}_{us}+V_{cb}V^{*}_{cs})\times(b_{3} - \frac{b_{3}^{ew}}{2})\right]
\label{K0p0}
\eea
where $r_{\chi}=\frac{2m_{K}^{2}}{(m_{b}-m_{d})(m_{d} + m_{s})}$
is the chirally enhanced terms in the penguin $O_{6}$ matrix
element. 
We also need the $B^{+}\to \pi^{+}\pi^{0} $ amplitude:
\bea
\kern -0.6cm &&A(B^{+}\to \pi^{+}\pi^{0}) = -i\frac{G_{F}}{2}f_{\pi}F^{B\pi}_{0}(m^{2}_{\pi})
(m_{B}^{2}-m_{\pi}^{2})\nonumber \\
&&\biggl(V_{ub}V^{*}_{ud}(a_{1} + a_{2})
+(V_{ub}V^{*}_{ud} +V_{cb}V^{*}_{cd})\nonumber \\
\kern -0.6cm&&\times \frac{3}{2}(a_{9}-a_{7}+ a_{10} +a_{8}r_{\chi})\biggr)
\label{p2p1}
\eea
We see that the $B\to K\pi$ decay
amplitudes consist of a QCD penguin(P) $a_{4} + a_{6}r_{\chi} $ ,
a color-allowed tree(T) $a_{1}$, a color-suppressed  tree(C) $a_{2}$
, a color-allowed electroweak penguin (EW) $a_{9}-a_{7}$, a 
color-suppressed  electroweak penguin (EWC) $a_{10}+ a_{8}r_{\chi}$ term.

Similar expressions for the QCD coefficients for  $B\to PV$ decays with 
hard-scattering corrections and annihilation terms used in the calculations
are not shown here, but can be found in  \cite{QCDF1,QCDF2,Zhu3,Alexan}.
For the CKM matrix elements, since the inclusive and exclusive data on
$|V_{ub}|$ differ by a large amount and  the higher inclusive  data
exceeds the unitarity limit for 
$R_{b} = |V_{ud}V_{ub}^{*}|/|V_{cd}V_{cb}^{*}| $ with the current value
$\sin(2\beta)=0.682\pm 0.019 $ \cite{PDG}, we shall determine $|V_{ub}|$ from
the more precise $|V_{cb}|$ data \cite{Kang}. As mentioned in \cite{Pham3}, we have~:
\be
\kern -0.5cm \vert V_{ub}\vert=  \frac{\vert V_{cb}V_{cd}^{*}\vert}{\vert V_{ud}^{*}\vert} \vert  \sin \beta 
\sqrt{1+\frac{\cos^2 \alpha}{\sin^2 \alpha}} .
\label{Vub}
\ee
 With $\alpha=(93.7\pm 10.6)^{\circ}$ \cite{Belle} and 
$\vert V_{cb}\vert = (41.78\pm 0.30\pm 0.08)\times 10^{-3}$
 \cite{Barberio}, we find, neglecting the errors,
\be
\vert V_{ub}\vert = 3.56\times 10^{-3}.
\label{Vub1}
\ee
in good agreement with the exclusive data in the range
$\vert V_{ub}\vert = 3.33-3.51$ \cite{Barberio} . A recent UT fit also gives 
$\vert V_{ub}\vert = (3.61\pm 0.12)\times 10^{-3} $  and 
$\vert V_{cb}\vert = (41.53\pm 0.30\pm 0.66)\times 10^{-3}$
close to the above values \cite{Bona}.
The measurements of the $B_{s}-\bar{B}_{s}$ mixing 
also allow the extraction of $|V_{td}|$ from $B_{d}-\bar{B}_{d}$
mixing data. The  current determination \cite{Abulencia} gives
$|V_{td}/V_{ts}|= (0.208^{+0.008}_{-0.006})$ which in turn can be used to
determined the  angle $\gamma$ from the unitarity relation \cite{CKM}:
\be
\kern -0.5cm \vert V_{td}\vert= \frac{\vert V_{cb}V_{cd}^{*}\vert}{\vert V_{tb}^{*}\vert} \vert  \sin \gamma
\sqrt{1+\frac{\cos^2 \alpha}{\sin^2 \alpha}}.
\label{Vtd}
\ee
with $|V_{tb}|=1 $, we find $\gamma = 67.6^{\circ} $
which implies an angle $\alpha = 90.7^{\circ}$, in good agreement  with
the  new Belle value $\alpha=(93.7\pm 10.6)^{\circ}$ \cite{Belle} 
mentioned above. The value $\gamma = 67.6^{\circ} $ is also consistent
with the current UT fit value $\gamma = (70.3\pm 3.7)^{\circ}$ \cite{Bona}.
In the following in our calculations, we shall
use the unitarity triangle values for $\vert V_{ub}\vert $ and $\gamma$. For 
other hadronic parameters we use the values in 
Table 1  of \cite{QCDF2} and take $m_{s}(\rm 2\,GeV)=80\,\rm MeV$.
For the $B\to \pi$ and $B\to K$ transition form factor, we use the current
light-cone sum rules central value \cite{Zwicky}~:
\be
F^{B\pi}_{0}(0)= 0.258,\quad   F^{BK}_{0}(0) = 0.33
\label{FBK}
\ee

\begin{table*}[ht]
\begin{center}
\begin{tabular}{|c|c|c|c|c|}
\hline
 Decay &$A\times 10^{8}{\rm GeV(QCDF)}$&${\rm BR\times 10^{6}( QCDF)}$&${\rm BR\times 10^{6}( exp.)}$\cite{PDG,HFAG} \\ 
\hline
\hline
$B^{+}\to \pi^{+}\pi^{0}$&$2.162- 1.112\,i$ &$5.535$ &$ 5.5 \pm 0.4$\\
$B^{+}\to \rho^{0}\pi^{+}$&$0.925- 2.752\,i$ &$7.732$ &$ 8.3 \pm 1.2$\\
$B^{+}\to \rho^{+}\pi^{0}$&$1.863- 3.055\,i$ &$11.744$ &$ 10.9 \pm 1.4$\\
\hline
$B^{+}\to K^{+}\pi^{0}$&$0.725+ 3.244\,i$ &$10.266$ &$ 12.94^{+0.52}_{-0.51}$\\
$B^{+}\to K^{0}\pi^{+}$&$0.162 + 4.399\,i$&$18.002$ & $ 23.79\pm 0.75$\\
$B^{0}\to K^{+}\pi^{-}$&$0.887+ 4.180\,i $ &$15.782$&$19.57^{+0.53}_{-0.52}$ \\
$B^{0}\to K^{0}\pi^{0} $&$-0.016- 2.817\,i$&$6.863$ &$9.9\pm 0.5$ \\
\hline
$B^{+}\to K^{+}\rho^{0}$&$1.422 + 0.4.483\,i$&$2.052$  &$3.7\pm 0.5 $\\
 $B^{+}\to K^{0}\rho^{+} $&$2.463 - 0.363\,i$&$5.637$&$8.0 \pm 1.5$\\
$B^{0}\to K^{+}\rho^{-}$ &$2.608 + 0.466\,i$&$5.943 $&$7.0\pm 0.9$\\
$B^{0}\to K^{0}\rho^{0} $ &$-2.164 + 0.411\,i$&$4.107 $&$4.7 \pm 0.6$\\
\hline
$B^{+}\to K^{*+}\pi^{0} $ &$-1.495 +0.786\,i$&$2.589 $&$8.2\pm 1.8$\\
$B^{+}\to K^{*0}\pi^{+} $ &$-1.876 -0.022\,i$& $3.206 $&$10.1^{+0.8}_{-0.9}$\\
$B^{0}\to K^{*+}\pi^{-}$&$-1.657 +0.946\,i$&$  3.084$&$ 8.4 \pm 0.8$\\
$B^{0}\to K^{*0}\pi^{0}$&$1.003 +0.128\,i$&$0.867$ & $ 3.3\pm 0.6$\\
\hline
\end{tabular}
\end{center}
\caption{ The measured and computed QCDF branching ratios shown with the QCDF amplitudes for $B\to PV$ decays }
\label{tab-res1}
\end{table*}

  The computed branching ratios  with $\rho_{A}=1$,
$\rho_{H}=1,\phi_{H}=0$ and $\phi_{A}=-55^{\circ}$ as in scenario S4 
of \cite{QCDF2} are shown in Table \ref{tab-res1}.
As can be seen,  QCDF with power corrections from penguin
annihilation as in S4 \cite{QCDF2,Zhu} could bring the branching ratios closer
to experiments. With a different choice of the annihilation parameters, as
given in  \cite{Cheng2}, one could  increase further the predicted decay rates
to values consistent with experiments. For the  CKM-allowed 
tree-dominated decays, as shown in Table \ref{tab-res1} and
in\cite{Cheng} , the predicted 
$B^{+}\to \pi^{+}\pi^{0},\rho^{0}\pi^{+}, \rho^{+}\pi^{0}$ decay rates
agree well with experiments. Therefore we can use the QCDF tree amplitude
for $\Delta S=1$ $B\to PP,PV$ in the determination of the phases of
the decay amplitudes. For this purpose, one needs to express the 
$\Delta S=1$ $B\to PP,PV$ decay amplitudes in terms of isospin amplitudes.
Following \cite{Deshpande,Isola},  we have, for $B\to K\pi$, in the 
notation of \cite{Isola}:
\bea
&&A_{K^+\pi^0} = 
 {2\over 3} B_3 + \sqrt{{1\over 3}} 
(A_1+B_1),   \nonumber\\
&&A_{ K^0\pi^+} =
{-\sqrt{2}\over 3} B_3 + \sqrt{{2\over 3}} 
(A_1+B_1),\nonumber\\
&&A_{K^+\pi^-} =
{\sqrt{2}\over 3} B_3 + \sqrt{{2\over 3}} 
(A_1-B_1),\ \ \nonumber \\
&&A_{ K^0\pi^0} = 
{2\over 3} B_3 - \sqrt{{1\over 3}} 
(A_1-B_1),
\label{ampl}
\eea
with $B_{1},B_{3}$ the  $I=1/2$ and $I=3/2$ isospin amplitudes in terms of
the decay amplitudes:
\bea
&&A_{1}=\frac{\sqrt{6}}{4}\,(A_{K^0\pi^+} +A_{K^+\pi^-})\nonumber\\
&&B_{1}=\frac{1}{\sqrt{3}}\,A_{K^+\pi^0} +\frac{\sqrt{6}}{12}\,A_{K^0\pi^+}-\frac{\sqrt{6}}{4}\,A_{K^+\pi^-}\nonumber\\
&& B_{3}=A_{ K^+\pi^0} -\frac{1}{\sqrt{2}}\,A_{K^0\pi^+}
\label{A1B1B3}
\eea
with the expressions in QCDF given by 
\bea
\kern -0.7cm && A_{1} = -i\frac{G_{F}}{2}f_{K}F^{B\pi}_{0}(m^{2}_{K})\frac{\sqrt{3}}{2}
(m_{B}^{2}-m_{\pi}^{2})\biggl(V_{ub}V^{*}_{us}\,a_{1}\nonumber \\
\kern-0.7cm&&+(V_{ub}V^{*}_{us}+V_{cb}V^{*}_{cs})[2\,a_{4}  +\frac{1}{2}a_{10}+ (2\,a_{6}+ \frac{a_{8}}{2})r_{\chi}]\biggr)
\nonumber \\ 
\kern -0.7cm&&-i\frac{G_{F}}{2}f_{B}f_{K}f_{\pi}\nonumber \\
\kern -0.7cm&&\times\biggl(V_{ub}V^{*}_{us}b_{2} + (V_{ub}V^{*}_{us}+V_{cb}V^{*}_{cs})\times(b_{3} + \frac{3}{2}b_{3}^{ew})\biggr)
\label{A1}
\eea
For $B_{1}$, we have:
\bea
\kern -0.7cm && B_{1} =  -i\frac{G_{F}}{2}f_{K}F^{B\pi}_{0}(m^{2}_{K})
(m_{B}^{2}-m_{\pi}^{2})\frac{\sqrt{3}}{2}\nonumber \\
\kern -0.7cm&&\left(V_{ub}V^{*}_{us}\frac{1}{3}a_{1}+(V_{ub}V^{*}_{us}+V_{cb}V^{*}_{cs})[\frac{1}{2}(a_{10}+a_{8}r_{\chi})]\right)\nonumber \\ 
\kern -0.7cm&& -i\frac{G_{F}}{2}f_{\pi}F^{BK}_{0}(m^{2}_{\pi})(m_{B}^{2}-m_{K}^{2})\nonumber \\
\kern -0.7cm&&\left(V_{ub}V^{*}_{us}\frac{2}{3}a_{2}+(V_{ub}V^{*}_{us}+V_{cb}V^{*}_{cs})\, {2}(a_{9}-a_{7})\right) \nonumber \\ 
\kern -0.7cm&&-i\frac{G_{F}}{2}f_{B}f_{K}f_{\pi} \nonumber \\ 
\kern -0.7cm&&-\biggl[V_{ub}V^{*}_{us}b_{2} + (V_{ub}V^{*}_{us}+V_{cb}V^{*}_{cs})
  b_{3}^{ew}\biggr] 
\label{B1}
\eea
and for $B_{3}$ 
\bea
\kern -1.0cm && B_{3} = -i\frac{G_{F}}{\sqrt{2}}f_{K}F^{B\pi}_{0}(m^{2}_{K})
(m_{B}^{2}-m_{\pi}^{2})\nonumber \\
\kern -1.0cm &&\left(V_{ub}V^{*}_{us}a_{1}
+(V_{ub}V^{*}_{us}+V_{cb}V^{*}_{cs})-\frac{3}{2}(a_{10}+a_{8}\,r_{\chi})\right) \nonumber \\ 
\kern -1.0cm && -i\frac{G_{F}}{\sqrt{2}}f_{\pi}F^{BK}_{0}(m^{2}_{\pi})(m_{B}^{2}-m_{K}^{2}) \nonumber\\ 
\kern -1.0cm &&\left(V_{ub}V^{*}_{us}a_{2}+(V_{ub}V^{*}_{us}+V_{cb}V^{*}_{cs})\times  \frac{3}{2}(a_{9}-a_{7})\right)\nonumber \\ 
\label{B3}
\eea
  We see that $B_{3}$ does not contain the strong penguin $a_{4}$ 
 and $a_{6}$ terms.  In the $SU(3)$ limit, apart from the small 
electroweak penguin terms, the main contribution to $B_{3}$ comes
 from the large color-favored
$(a_{1} + a_{2})$ term, as in   $B^{+}\to \pi^{+}\pi^{0}$ decay, for
  which QCDF without   the strong penguin contributions, is quite reliable,
as can be seen from the good agreement with experiments for 
 $B^{+}\to \pi^{+}\pi^{0},\rho^{0}\pi^{+}, \rho^{+}\pi^{0}$ decays
shown in  Table \ref{tab-res1}. The relation between $B_{3}$ and the 
$B^{+}\to \pi^{+}\pi^{0}$ decay amplitude can also be obtained in a general
proof based on a model-independent  approach to charmless $B\to PP$ 
decays,  given recently in \cite{He}. In terms of the 
$SU(3)/U(3)$ invariant amplitudes, one has, putting aside  the CKM factor:
\bea
\kern -1.0cm && T^{B_{u}}_{\pi^{-}\pi^{0}} = \frac{8}{\sqrt{2}}\,C^{T}_{\bar{15}} \nonumber\\
\kern -1.0cm && T^{B_{u}}_{\pi^{0}K^{-}}=\frac{1}{\sqrt{2}}\,(C^{T}_{\bar{3}}
-C^{T}_{\bar{6}}+ 3\,A^{T}_{\bar{15}} + 7\,A^{T}_{\bar{15}}) \nonumber\\
\kern -1.0cm && T^{B_{u}}_{\pi^{-}\bar{K}^{0}}=(C^{T}_{\bar{3}}
-C^{T}_{6}+ 3\,A^{T}_{\bar{15}} -\,C^{T}_{\bar{15}})
\label{SU3}
\eea
From Eq. (\ref{SU3}), we get:
\be
B_{3}= \sqrt{2}\,T^{B_{u}}_{\pi^{0}K^{-}} - T^{B_{u}}_{\pi^{-}\bar{K}^{0}}=
\frac{8}{\sqrt{2}}\,C^{T}_{\bar{15}} = T^{B_{u}}_{\pi^{-}\pi^{0}}
\label{B3SU3}
\ee
in agreement with QCDF in the $SU(3)$ limit. This relation can also be
derived in a simple manner by using the topological amplitudes. We
 have\cite{Cheng}~:
\bea
&&A_{K^+\pi^0} = -\frac{1}{\sqrt{2}}(p^{\prime} + t^{\prime} + c^{\prime}),
 A_{ K^0\pi^+} = p^{\prime}  \nonumber\\
&&A_{ \pi^{+}\pi^0} = -\frac{1}{\sqrt{2}}( t + c) \nonumber\\
&&B_{3} =  -(t^{\prime} + c^{\prime})
\label{CT}
\eea
showing  $B_{3}=\sqrt{2}\,A_{\pi^+\pi^0}$ in the $SU(3)$ limit. 

 Given QCDF for the CKM-favored tree-dominated decay amplitudes, 
the $SU(3)$ breaking effects   can be automatically
taken into account in the QCDF expressions for penguin-dominated decays. The
point we made in  this paper is that QCDF works well for processes with
large color-favored tree contribution, but without the strong penguin
terms. The agreement with experiments for 
 $B^{+}\to \pi^{+}\pi^{0},\rho^{0}\pi^{+}, \rho^{+}\pi^{0}$ 
measured branching ratios and the rather well-known short-distance Wilson 
coefficients for the tree operator shows that the central values for the
form factors and decay constants involved  are consistent with 
experiments and can be used in QCDF calculations with  penguin-dominated
decays. Thus the uncertainties for the QCDF branching ratios depend only 
on the accuracy of the  measured
$B^{+}\to \pi^{+}\pi^{0},\rho^{0}\pi^{+},\rho^{+}\pi^{0}$  branching
ratios, which are  $10\%$ while the theoretical errors and 
uncertainties  in the current QCDF calculations  are quite
large\cite{QCDF1,QCDF2,Cheng}. This shows the  advantage of using
the measured $B^{+}\to \pi^{+}\pi^{0},\rho^{0}\pi^{+},\rho^{+}\pi^{0}$ 
branching ratios to obtain the correct form factor values
for  QCDF calculations of the   $B\to K\pi$,$K\rho$, $ K^{*}\pi$ decay 
rates and in particular for the $I=3/2$  isospin amplitude $B_{3}$,
though the $SU(3)$ relation between $B_{3}$ and the $B^{+}\to \pi^{+}\pi^{0}$ 
amplitude in Eq. (\ref{B3SU3}) or Eq. (\ref{CT}) is useful for a qualitative
argument that $B_{3}$  is exactly the $B^{+}\to \pi^{+}\pi^{0}$ $I=2$
amplitude in the  $SU(3)$ limit. For the  penguin-dominated decays, we 
do not expect QCDF to  produce a correct penguin amplitude
in the $B\to K\pi$,$K\rho$, $K^{*}\pi$ decays which could have power
correction terms like the penguin annihilation mentioned in the
literature\cite{QCDF2,Zhu,Cheng2}, especially for the  predicted
$K^{*}\pi$  branching ratios which are  below the measured values
 by more that $30\%$.
\vspace*{-0.50cm}
\section{DETERMINATION OF PHASES OF THE $\Delta S=1$ $B\to PP,PV$ DECAY AMPLITUDES}
With the  $I=3/2$ amplitude  given by QCDF, we now proceed to the
determination of the relative phase between the tree and penguin amplitudes.
 
 As shown in  \cite{Isola}, by taking the sum of the $B^{+}$ and $B^{0}$
 absolute square of the amplitudes  $|A|^{2}$ or  the decay rates, from
Eqs. (\ref{ampl}), we have~: 
\be
\kern -1.0cm |A_1+B_1|^{2} = |A_{K^+\pi^0}|^{2} + |A_{ K^0\pi^+}|^{2}
- {2\over 3} |B_3|^{2}   
\label{AB1}
\ee
\be
\kern -1.0cm |A_1-B_1|^{2}=|A_{K^+\pi^-}|^{2} + |A_{ K^0\pi^0}|^{2} -
{2\over 3} |B_3|^{2} . 
\label{AB2}
\ee
With the lengths of the sides $A_{1} + B_{1}$ and $A_{1} - B_{1}$ given by the
decay rates of the four $B\to K\pi$ decay modes in
Eqs. (\ref{AB1}-\ref{AB2}) , the  angles of the triangle formed with
the decay amplitude,  $B_{3}$, and  with $A_{1} + B_{1}$
and $A_{1} - B_{1}$ , respectively . This gives us the 
relative phase between the $ I=3/2$ 
 tree and the $ I=1/2$ penguin amplitudes for a precise 
test of the QCDF. Clearly, isospin amplitudes are needed 
to obtain  the phases in $B\to K\pi$,$K\rho$, and $K^{*}\pi$ decays 
which are in the interference term between $B_{3}$ and 
 $A_{1} + B_{1}$ and between $B_{3}$ and  $A_{1} - B_{1}$, and each of the
length  $|A_{1} + B_{1}|$ and $|A_{1} - B_{1}|$ depends on 
the branching ratios of two decay modes.
 Let $\delta_{1,2}$ be the relative phase  between 
$B_{3}$ and $A_{1} + B_{1}$, and between $B_{3}$ and $A_{1} - B_{1}$
respectively, from  Eqs. (\ref{ampl}) and using 
Eqs. (\ref{AB1}-\ref{AB2}), we have:
\be
\kern -1.0cm \cos(\delta_{1})
=\frac{\sqrt{3}(2\,|A_{K^+\pi^0}|^{2} - |A_{ K^0\pi^+}|^{2}-
  |B_{3}|^{2}/3)}{4\,|B_{3}||A_1+B_1|}  
\label{delta1}
\ee
\be
\kern -1.0cm \cos(\delta_{2})
=\frac{\sqrt{3}(|A_{ K^+\pi^-}|^{2} - 2\,|A_{K^0\pi^0}|^{2}+ |B_{3}|^{2}/3)}{4\,|B_{3}||A_1-B_1|} 
\label{delta2}
\ee
 Since all the four penguin-dominated decay modes have similar decay rates, 
the  differences $|A_{ K^0\pi^+}|^{2} -2\,|A_{K^+\pi^0}|^{2}$
and  $|A_{ K^+\pi^-}|^{2} - 2\,|A_{K^0\pi^0}|^{2}$ become small,
errors and uncertainties in the measured decay rates would make it  
difficult to obtain a correct  value for $\cos(\delta_{1})$ 
and $\cos(\delta_{2})$. Another problem which could affect the phase 
determination is the consistency of the four measured decay rates 
imposed on  by an  isospin relation between the  decay rates which is
 given  as\cite{Isola,Lipkin}, with QCDF values for $|B_{3}|^{2}$  and ${\rm Re} (B_3^* B_1)$ :
\bea
&&\kern -1.0cm |A_{K^+\pi^-}|^{2} -2\,|A_{K^0\pi^0}|^{2}=\nonumber\\
 &&\kern 1.0cm-\biggl[|A_{K^0\pi^+}|^{2}  -2\|A_{ K^+\pi^0}|^{2}\biggr]\nonumber\\
&&\kern 1.0cm-\biggl[{4\over 3}|B_3|^2+{8\over {\sqrt 3}}{\rm Re} (B_3^* B_1)\biggr]_{K\pi}\,\,
\label{D2}
\eea
\begin{table*}[ht]
\begin{center}
\begin{tabular}{|c|c|c|c|c|}
\hline
 Decay
 &$\delta_{1}{\rm (deg)(QCDF)}$&$\delta_{1}{\rm (deg)(exp.)}$&$\delta_{2}{\rm (deg)(QCDF)}$&$\delta_{2}{\rm (deg)(exp.)}$\\ 
\hline
\hline
$B\to K\pi$&$71.891$ &$77.296\pm 15$ &$ 68.968$&$ 75.199\pm 15$(\rm estimated)\\
\hline
$B\to K\rho$&$113.701$&$109.217$(\rm estimated) &$110.925$&$110.638 $\\
\hline 
$B\to K^{*}\pi$&$67.838$&$73.351$ (\rm estimated)&$58.194 $&$68.078$\\
\hline
\end{tabular}
\end{center}
\caption{ The relative isospin phases given by QCDF and obtained from
the measured decay rates for   $B\to K\pi$,$K\rho$ and  $ K^{*}\pi$  
decays. The numbers  marked as ``estimated'' are the phases obtained 
with isospin relation as explained in the text.
Errors are estimated to be in the range $\pm (10-15)^{\circ}$}
\label{tab-res2}
\end{table*}
This relation gives  a branching ratio $8.98\times 10^{-6}$ for
$B^{0}\to K^{0}\pi^{0}$  to be cmpared with the  measured value of 
$(9.93\pm 0.49) \times 10^{-6}$ which produces a cancellation in the
quantity $ |A_{ K^+\pi^-}|^{2} - 2\,|A_{K^0\pi^0}|^{2}$ in Eq. (\ref{delta2})
and a phase $\delta_{2}$ near $90^{\circ}$, which deviates largely from
the  phase between $B_{3}$ and $A_{1} + B_{1}$, in 
contradiction with isospin analysis, since $|B_{1}|$ is small compared with
$|A_{1} + B_{1}|$ and $|A_{1} - B_{1}|$, the difference $\delta_{2}-\delta_{1}$
should be small. Using the above estimated branching ratio for
$B^{0}\to K^{0}\pi^{0}$,  one would obtain 
 $\delta_{2}= 75.199^{\circ}$, close to the value $77.296^{\circ}$
for $\delta_{1}$, consistent with isospin analysis. Thus a correct
value for $\delta_{2}$ consistent with $\delta_{1}$ requires  a lower 
 value for  $B^{0}\to K^{0}\pi^{0}$ branching ratio . This lower value for
$B^{0}\to K^{0}\pi^{0}$ could turn out to be the correct value, as 
over the years, the $B^{0}\to K^{0}\pi^{0}$ branching ratio has
decreased  to the present value.

The phases  for $B\to K\rho$ and  $ B\to K^{*}\pi$ decays  can be 
obtained from the above expressions by making a  straightforward 
substitution with the $K\rho$ and  $ K^{*}\pi$ decay rates. In 
Table \ref{tab-res2}, we give the relative isospin 
phases $\delta_{1,2} $ for  $B\to K\pi$,$K\rho$, and  $ K^{*}\pi$
obtained from  QCDF and from  the measured decay rates.

  As with $B\to K\pi$ decays, the  determination of $\delta_{1} $ in
$B\to K\rho$ decays  is also subject to  large uncertainties, with 
almost a cancellation in the difference
($|A_{K^0\rho^+}|^{2}  -2\,|A_{ K^+\rho^0}|^{2}$) , one would  get a
 value $\delta_{1}=98.791^{\circ} $, very different from the
value  $110.638^{\circ}$ for $\delta_{2}$. In fact, using the isospin
relation for $K\rho$ given as~:
\bea
&&\kern -1.0cm(|A_{K^0\rho^+}|^{2} -2\,|A_{K^+\rho^0}|^{2})=\nonumber\\
 &&\kern 1.0cm -(|A_{K^+\rho^-}|^{2}  -2\|A_{ K^0\rho^0}|^{2})\nonumber\\
&&\kern 1.0cm+\biggl[-{4\over 3}|B_3|^2-{8\over {\sqrt 3}}{\rm Re} (B_3^*B_1 )\biggr]_{K\rho}\,\,
\label{D1}
\eea
we would get  a branching ratio 
$(9.15\pm 1.2) \times 10^{-6}$ for $B^{+}\to K^{0}\rho^{+}$ 
higher  than the measured value  of $(8.0^{+1.5}_{-1.4}) \times 10^{-6}$.
This predicted branching ratio then gives $\delta_{1}=109.217^{\circ} $ 
close to the value  $110.638^{\circ} $  for $\delta_{2}$, consistent
with the fact that,  as in $B\to K\pi$  decays, since $|B_{1}|$ is small 
compared  with the penguin amplitude $|A_{1}|$,  $\delta_{1}$ and 
 $\delta_{2}$ should be close to each other, as seen from the QCDF
values given in  the Table (\ref{tab-res2}).

  Similar  problem also appears in  $ B\to K^{*}\pi$ decay, as the 
isospin relation similar to that for $ B\to K\rho$ in Eq. (\ref{D1})
 would give a branching ratio  $(6.3\pm 2.2)\times 10^{-6}$ for
$B\to K^{*}\pi$ decay, lower than  the measured value of 
$(8.2\pm 1.9)\times 10^{-6}$. For  this reason,  the phases
 $\delta_{1,2}$  for $B\to K^{*}\pi$ decay are obtained using only
 the  $B^{0}\to K^{*+}\pi^{-}$ and $B^{0}\to K^{*0}\pi^{0}$  decay
 rates and the isospin relation, as   shown in  Table (\ref{tab-res2}).
We note that for $B\to K\pi$ decays, the errors on the phases 
$\delta_{1,2}$, are around $15^{\circ} $. This could be due to the large
cancellation bertween the measured branching ratios which however,
have small errors, of the order few percent. For this reason we will not
give errors on the phases for the $K\rho$ and  $ K^{*}\pi$ decays for which
the errors are more than 10 percent. We note also that the
errors for $B\to K\pi$ shown in the Table (\ref{tab-res2}),  is comparable to
the errors found in the determination of the relative phase between the
three-gluon and the one photon annihilation amplitudes of the $\psi(2S)$
decays to pseudoscalar meson pairs, for which a relative phase of
$(-82\pm 29)^{\circ}$ or $(+121\pm 27)^{\circ}$ is found in \cite{Bai}.
What is remarkable with the result we found is that, all  the phases for
$B\to K\pi$,$K\rho$,  and $ K^{*}\pi$ decays  obtained with the central
values for the measured branching ratios consistently show  only small
deviations  from the  QCDF values. The implication of this result is that
one may need power correction terms, probably of perturbative QCD
origin, to bring QCDF values close to the measured decay rates, without 
the need for a strong phase from  long-distance  rescattering effects.
\vspace*{-0.50cm}
\section{CONCLUSION}
\vspace*{-0.50cm}
  With the tree amplitude known from the QCDF tree-dominated $B\to PP,PV $
decays, we are able to determine the relative phases of the tree-penguin
interference term in  $B\to K\pi$,$K\rho$, and $ K^{*}\pi$ decays.  We find
that the phases in the tree-penguin interference terms differs slightly 
from the QCDF phases, in particular, with an uncertanity 
  $\pm 15^{\circ}$ more or less for $B\to K\pi$, Also for 
$K\rho$ and $ K^{*}\pi$ decays,  this uncertanity could be reduced
considerably with more precise data with LHCb and the coming 
super Belle. This would allow a precise test of QCDF.

\end{document}